# Investing in nature: Stakeholder's willingness to pay for Tunisian forest services

Islem Saadaoui*

***Abstract***

*This study explores the economic value of Aleppo pine forests, a unique and threatened ecosystem in the border region of central Tunisia. These forests play a vital role in supporting small rural communities, but face increasing pressures and restrictions on their use. This research aims to assign a monetary value to forest conservation, considering the region's specific socio-economic context. Strategies for empowering local residents as key actors in developing sustainable cross-border initiatives are further investigated. Employing the Contingent Valuation Method, a survey of 350 local residents and international users was conducted to assess their Willingness to Pay for forest conservation efforts. Logistic regression analysis revealed that sociodemographic factors, such as monthly income and preferred payment method, significantly influence both and the likelihood of participation. These findings highlight the feasibility and importance of reconciling economic development with ecological sustainability in this critical region.*

## 1. Introduction

Natural ecosystems generate goods and services that are essential to society's well-being (Haines-Young and Potschin, 2010). However, most of these products are exploited outside the market, without user fees for the population (Cenamor and Frishammar, 2021; Spangenberg and Settele, 2010; Farley and Costanza, 2010). Consequently, the absence of monetary value does not reflect consumers' willingness to pay for ecosystem and landscape preservation (Tian *et al*., 2020; Martín-López *et al*., 2007). This lack of economic value has led to the allocation of zero value to ecological services in decision-making (Mandle *et al*., 2021; Ouyang *et al*., 2020), resulting in the over-exploitation of ecosystems worldwide.

The study region, the Aleppo pine forests of central Tunisia, has suffered from overexploitation of its natural and environmental resources (Saadaoui *et al*.,2018), sometimes associated with irreversible degradation of vegetation cover caused by human and/or natural actions (Islem *et al*., 2014; Saadaoui, 2016).

These actions seriously threaten the ecological balance necessary for the sustainability of ecosystems and natural landscapes in Tunisia (Hammami *et al*., 2023; Jdaidi *et al*., 2023; Taghouti *et al*., 2021; Khalfaoui *et al*., 2020; Hasnaoui and Krott, 2019; Achour *et al*., 2018; Daly-Hassen *et al*., 2017; Daly-Hassen and Croitoru, 2013; Campos *et al*., 2008). The characteristics and components of these landscapes, as well as socio-economic evolution and the rapid urban sprawl of recent decades, explain these phenomena. To overcome these problems, it is essential to develop appropriate tools to preserve the vital resources to sustainable development in the study region. Appropriate procedures for valuing environmental assets as "gifts of nature" are needed to raise awareness of the need to preserve them.

---

* Université du Littoral Côte d'Opale - ULR 4477 - TVES - Territoires Villes Environnement & Société, Dunkerque, France.
Corresponding author: islem.saadaoui@univ-littoral.fr

In general, there is no conventional market for this type of environmental good. Consequently, it is imperative to set up procedures that take into account the specific characteristics of forest and mountain landscapes. Forests represent a first-rate natural and environmental resources (Brockerhoff *et al.*, 2017; Mori *et al.*, 2017; Dobbs *et al.*, 2011), offering a variety of goods and services that can be grouped into three main categories:

Production of traditional goods with a value derived directly from the market: This category includes everyday forest products such as wood, as well as non-wood products such as mushrooms, honey and medicinal plants.

Environmental services with a use value that can be estimated using methods based on market-revealed preferences: these include soil protection, erosion control, reservoir silting prevention and recreational activities.

The total economic value (TEV) of an ecosystem encompasses all the benefits people derive from it, both directly (use values) and indirectly (passive values) (Davidson, 2013). Environmental services with no direct use value, but characterized by non-use values (option values, existence values, bequest values) assessed using non-conventional approaches based on stated values (Contingent Valuation Methods): Examples of these services include biodiversity, pollution reduction, etc. (Mori *et al.*, 2017; Chazdon *et al.*, 2016).

This is the background to this study. Local authorities frequently express the need to better understand user preferences. What landscape attributes are they looking for? What is the users willingness to pay for the various attributes of a natural or recreational landscape? When a territorial project can have an impact on the environment, what are the criteria for valuing it, especially in a context of budgetary constraints, especially when the project does not involve marketable goods?

## 2. Methodology

Assessing the value of landscape attributes in natural recreational areas is a challenge due to the absence of a market with defined prices. Consequently, it is necessary to devise a method for assigning values to the landscape attributes of selected resorts, values that could be used in public decision-making. With this in mind, various preference assessment methods have been devised and developed to compensate for the absence of a market.

Two categories of methods can be distinguished: on the one hand, direct methods that simulate a market for non-market goods using one or more hypothetical scenarios (such as contingent valuation or choice experiments); on the other hand, those that reveal individuals' preferences by observing their behavior on complementary (such as hedonic prices, travel costs) or substitute (such as protection costs) markets (Oueslati *et al.*, 2008).

The contingent valuation method (CVM) has long been a popular tool for valuing non-market environmental assets. Recent studies continue to confirm its effectiveness (Manero *et al.*, 2024; Grazhdani, 2024; Blasi *et al.*, 2023; Baymumi-nova *et al.*, 2023; Raihan, 2023; Manero *et al.*, 2022; Viti *et al.*, 2022; Perni *et al.*, 2021; Cuccia, 2020; Guijarro and Tsinaslanidis, 2020; Rogers *et al.*, 2019; Dupras *et al.*, 2018; Krause *et al.*, 2017; del Saz-Salazar and Guaita, 2013), building upon a strong foundation established in earlier research (Randall *et al.*, 1983; Durden and Shogren, 1988; d'Arge and Shogren, 1989; Milne, 1991; Bateman and Turner, 1992; Jones, 1997; Oglethorpe and Miliadou, 2000)

It involves a direct survey of park users, asking them about their willingness to pay (WTP) based on hypothetical scenarios. Each scenario represents a change in one or more attributes, with an associated means of payment.

The integration of biodiversity conservation into development plans for cross-border mountain areas calls on a variety of concepts and processes, most often combining considerations relating to the ecosystem, nature, heritage-identity and economic interest of the environments (Saadaoui *et al.*, 2018).

This study proposes applying the contingent valuation method based on carrying out a field survey to determine the price that each person would be prepared to pay, in other words their willingness to pay for the conservation of an environmental asset. This study aims to conserve the nature of a small forest that has been unused for around a century, and to develop a natural recreational area in the cross-border mountains. The contingent valuation method is being tested at two sites: Dernaya forest (unexploited forest), an area of fallow land and forest clearings (Bouchebka). The research aims to identify two dinstinct willingness-to-pay values: the first is a conservation value (CV) and the second a management value (MV).

The criteria defined for CV and MV are determined using data collected from inhabitants of cross-border areas. The contingent valuation method is applied to a sample of 300 households from three regions in the mountain areas and 50 Algerian passenger respondents from the Bouchebka region (in Feriana).

## 2.1. Contingent Valuation Method

Contingent Valuation Method (CVM) is a survey-based technique used to estimate the value people place on environmental resources that lack a traditional market. People are asked their willingness to pay (WTP) to avoid degradation or improve these services.

How it Works: CVM creates a hypothetical market scenario where respondents are presented with a specific change in the ecosystem's condition and asked their WTP for that change. The researcher and respondent engage in a dialogue (through a well-designed questionnaire) to understand the respondent's true value for the ecosystem.

Consumer theory suggests that people make choices to maximize their satisfaction (utility). This study leverages this theory by using CVM to assign an economic value to environmental assets based on people's WTP. Essentially, the survey creates a hypothetical market to capture subjective values and estimate an objective value that could inform pricing decisions for environmental conservation or development projects. This section will detail the survey design and questionnaire structure used for data collection.

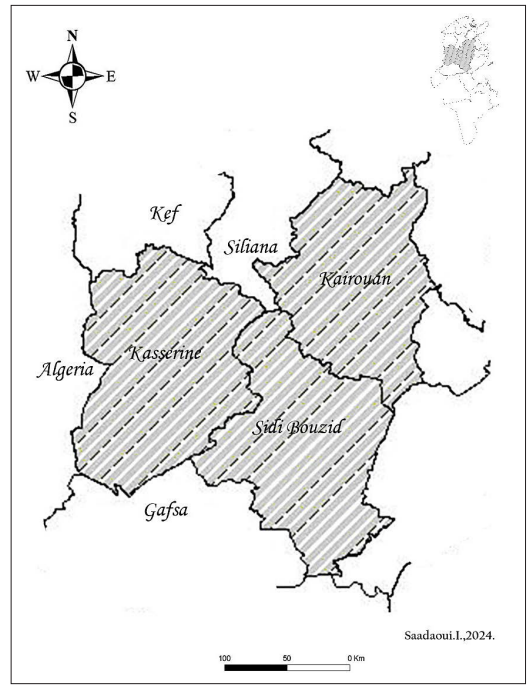

Figure 1 - Location map of the study site.

### 2.1.1. Site selection

Location: This study investigates two sites in the Kasserine governorate on the Tunisian-Algerian border: Allepo pine Forest of Dernaya and the Bouchebka Allepo pine forest clearings (Figure 1).

Selection Criteria: Several factors influenced site selection. Firstly, we chose the central Tunisia due to its proximity to the Algerian market (Bouchebka being the main border crossing). Secondly, we aimed for consistency in influencing factors by selecting sites with similar natural features (leisure parks, hiking trails, unexploited forests). To ensure variety in landscapes, the study sites were chosen for their contrasting aesthetic appeal (family-friendly atmosphere, diverse leisure areas, cultural heritage, etc.). Both locations are situated in the Feriana region, which is easily accessible to the target population for the proposed project.

### 2.1.2. Assumptions and scenario design

With the study scope defined, we propose two hypotheses to guide the research. Understanding people's payment behavior requires assumptions about the explanatory power of certain variables. One hypothesis aligns with the theory of citizen well-being:

Hypothesis 1: Users recognize the need to contribute financially to support a sustainable development project in the study area.

Respondents' express payment intentions based on the specific scenario presented. Therefore, a clear and well-defined scenario is crucial. Offering the option to propose fixed prices or payment ranges significantly reduced the number of refusal-to-pay responses.

*Hypothesis adopted during the work carried out:* The analysis considers a sample of N indexed individuals i = 1, ..., N. Analysis focuses on whether one of the two projects was implemented for each individual in the sample, and note yi the coded variable associated with the project in question. Let $\forall i \in [1, N]$:
yi = (1.0) if the project has been carried out for individual i, and if the project has not been carried out for individual i; yi = (1.1).

Note here the choice of the (0, 1) coding commonly used for dichotomous models.

This allows us to define the probability of the event occurring as the expectation of the coded variable yi, since:

$$E(y_i) = Prob(y_i = 1) \times 1 + Prob(y_i = 0) \times 0 = Prob(y_i = 1) = p_i$$

The aim of the dichotomous models is to explain the occurrence of the envisaged event as a function of a certain number of characteristics observed for the respondents (individuals in the sample). By applying these models, we seek to specify the probability of achieving the objective of our study (the conservation and/or development project).

*Contingent valuation scenarios*

The researcher plays a key role in creating a believable scenario for the CVM survey. This scenario should clearly describe:

The good (environmental resource): What is it? What benefits does it provide? What actions are needed to improve or conserve it?

Payment method: How will respondents' contributions be collected (mandatory fees, entrance fees, voluntary donations, etc.)? Who will benefit from these contributions?

## 2.2. *Questionnaire design and administration*

### 2.2.1. *Developing the Questionnaire*

The CVM method requires a well-designed questionnaire to collect data on respondents' WTP. Our questionnaire addressed response formats, potential for non-responses, and potential biases. It included 15 questions.

### 2.2.2. *Targeted Sampling*

To enhance the credibility of our results and ensure participation, we employed targeted sampling. Since 60% of Feriana's population is employed (6,162 households), we focused on this group to target individuals likely to contribute financially. The survey was administered to 300 households and 50 Algerian tourists, with each household including at least one employed person.

### 2.2.3. *Scenario Presentation*

After socio-economic questions, respondents were presented with two options: Contribute annually to a fund via a tax or making voluntary quarterly donations. Before choosing a payment method, respondents selected one of two projects: *Maintain a forested area and develop a natural leisure park (with images provided). *Preserve nature in the Bouchebka mountains without development.

### 2.2.4. *Information collected*

The questionnaire gathered four types of important information for understanding factors influencing WTP decisions:

Site Use: Respondents' frequency of visiting or using the site.

Resource Assessment: Their perception of the natural resources' value.

Project Opinions: Their views on the proposed programs.

Sociodemographic Data: Information like age, education, and income.

This data helped us propose payment amounts relevant to respondents' perceptions.

Tourists: The same questionnaire was presented to Algerian tourists to gauge their interest in contributing to a Bouchebka forest leisure park project.

Payment Preferences: For residents, the goal

was to determine their preference between the two scenarios (conservation vs. development) and calculate an average "Consent to Pay" (CTP) value. For tourists, the CTP represented a potential park entrance fee.

Questionnaire Approach: A closed-ended survey design with dichotomous response options was employed (accept/reject) regarding offered payment amounts.

### 2.2.5. *Additional Information*

The questionnaire aimed to capture the following:

Resident Behavior: Understanding how residents interact with the ecological services provided by the natural environment.

Willingness to Pay: Assessing residents' WTP for preserving or enhancing the region's economic/environmental value.

Human-Environment Connection: Exploring the link between respondents and their natural surroundings.

Socio-economic data (gender, age, marital status, education, income) was also collected to support further analysis.

### 2.3. *Econometric models: Probit and Logit models*

The econometric treatment of willingness-to-pay is an important step in contingent valuation studies. Probit and Logit models are used to explain the values of Y through X, i.e. to estimate the probability that $Y_i = 1$ knowing $X_i$. Note that:

$$Pr(Y_i = 1/X_i) = Pr(X_i \theta + \varepsilon_i \geq 0 / X_i) = Pr(X_i \theta \geq - \varepsilon_i / X_i) = F\text{-}\varepsilon(X_i \theta).$$

The only difference between the Probit and Logit models is the specification of F. In both cases, the distribution of residuals is symmetrical, so F-ε can be replaced by Fε.

The Probit model corresponds to the Gaussian specification introduced in the previous section. F is therefore the distribution function of a center-reduced Gaussian, usually denoted Φ:

$$F(X_i\theta) = \Phi(X_i\theta) = \int X_i \theta \, [(e\text{-}t^2/2)/(\sqrt{2\pi})] \times 2t$$

The corresponding density, usually denoted ϕ, is:

$$f(X_i\theta) = \phi(X_i\theta) = e - (x_i \theta)^2/2/\sqrt{2\pi}$$

There is virtually no difference between these two laws, the introduction of the logistic law being simply motivated by its simplicity in this framework (Crépet *et al.*, 2009).

Hanemann's (1984) framework for discrete choice models based on a random utility function is adopted to implement the Logit model in this research, with the aim of converting responses (yes/no) to the question of whether to pay consent.

The respondent's utility function is given by u (j, y, x), which breaks down into two parts:

$$u(j, y, x) = n(j, y, x) + e_j \text{ avec } n(j, y, x)$$

It is important to consider that:
- ej: is the random part of the utility function, representing an unobservable component.
- y: represents the respondent's income.
- x: represents the individual characteristics on which the utility function depends.
- j: indicator variable equal to 1 if the respondent agrees to pay the proposed value and 0 if not.

The respondent agrees to pay an amount M, to conserve and/or improve a natural environment and preserve the region's natural heritage if:

$$[n(1, y - M, x) + e_1] > [n(0, y, x) + e_0].$$

Consequently, the respondent will agree to pay if the well-being is achieved or the condition of the good is improved and will refuse to pay the amount if the opposite is the case.

The formula for the probability of agreeing to pay is as follows:

$$Prob = Prob\,[n(1, y - M, x) + e_1 > n(0, y, x) + e_0]$$

The following form can be used:

$$Prob = Prob\,[n(1, y - M, x) - n(0, y, x) > e_0 - e_1]$$

Where:
$\eta = (e_0 - e_1)$ and $\Delta n = n(1, y - M, x) - n(0, y, x)$

The answer is then a random variable composed of a probability density.

The cumulative distribution function ($F\eta(\Delta n)$) can be thought of as the probability of accepting the amount M either:

$$Prob(\text{accept } M) = P_i = F\eta(\Delta n) = 1 - G\eta(\Delta n)$$

- Pi: the probability that respondent i agrees to pay the proposed amount and

- G η (∆ n) the complement: the probability of refusing this amount.

At this point, the question arises of the functional form of the individual utility function and the distribution function, for which specific models are used: the logit model and the linear regression model. In the logit model, it is a distribution function of a logistic variable, while in the linear regression model, Fη (∆ n) represents the distribution function of a normal distribution (Maddala et Lahiri, 1992).

According to Abichou and Zaibet (2008), two features make the logistic function interesting for modeling discrete choices: its interval is limited to 0 - 1. The logit function can be used as a probabilistic function, offering the possibility of being linearized by logarithmic transformation. If ∆ n follows a logistic distribution, then the cumulative distribution function is given by (Maddala, 1983):

$$Pi = F\eta (\Delta n) = \exp \Delta n / (1 + \exp \Delta n)$$

This leads to the following equation:

$$\log [F\eta (\Delta n) / (1 - F\eta(\Delta n))] = \Delta n$$

The logit model is thus written as follows:

$$\text{logit} = \log (Pi / (1 - Pi)) = \Delta n = WTP$$

## 3. Results

This section presents the findings from our survey of 300 households in the Feriana region and 50 Algerian tourists.

### 3.1. *Socioeconomic characteristics*

The majority of Feriana households were male-headed (92%) with young adults under 50 years old comprising 71.3% of respondents. Most heads of household had a high level of education (26% with higher education) and the average household size was 5 people (58% with more than 5 members). Dominant professions included agriculture, construction, and other manual labor (82%), with 34% earning between 200-400 €. Regarding the environment, 100% of respondents were aware of the surrounding forests. The primary activities reported were collecting forest products (56%) and grazing (37%). Despite proximity, 40% of respondents had not visited the mountainous areas.

Among the 50 Algerian passengers surveyed, 84% were men, 60% were under 35, and 34% had higher education. Most traveled in groups of two (58%) and stayed less than five days in the region (64%). The main reason for short stays was the lack of high-quality leisure facilities and services, with 75% viewing Feriana as a transit area.

### 3.2. *Willingness to pay propensity*

This study investigates the relationship between socioeconomic characteristics and residents' payment preferences towards a potential project, hypothesizing that these characteristics influence financial support decisions. Our results showed that 42% of respondents favored a leisure project, while 24% preferred a nature conservation project. Overall, 67% of respondents agreed to contribute financially to their chosen project.

Analyzing the independent variables revealed that income level, education level, household size, and prior forest experience influenced willingness to pay. Individuals with higher incomes and education were more likely to contribute, while willingness to pay decreased with household size. Those who had visited the forest environment were also more inclined to contribute financially. Interestingly, the activity undertaken in the forest mattered, with walkers demonstrating a greater willingness to pay compared to hunters. Finally, indecisiveness regarding project choice before payment requests was linked to a higher refusal rate (55%). These findings support the hypothesis that socio-economic characteristics play a significant role in financial contribution decisions, validating the initial hypothesis. Furthermore, the analysis of factors influencing WTP aligns with respondents' declarations, contributing to the economic evaluation's theoretical validity.

### 3.3. *Revealed Willingness to Pay*

The survey revealed a total annual WTP of € 6,595 for Feriana residents, with an average of € 32.7 per person. Participants favored both the leisure park development and natural environment conservation projects. While the development value was higher than the conservation value, the difference was not statistically significant. Residents were offered two payment methods (tax or voluntary contribution), which may influence the final average CTP amount. Those opting for voluntary payments represented 63.1% of the sample, but their individual contributions were less than half those who chose the annual tax. Income level significantly influenced the amount paid, with a clear increase observed across income brackets. Individuals in the highest income class (≥€ 800,000/month) contributed more than double the average value of other income classes.

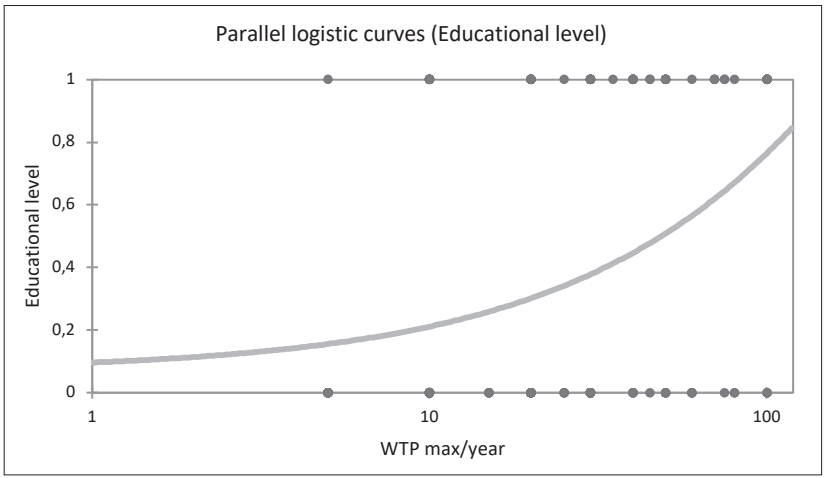

Figure 2 - Effect of the 'educational level' variable on the value of WTP.

For Algerian tourists, WTP for park entrance fees totaled € 272,500 for 50 visitors, translating to an average of € 5,450 per person. This value may vary depending on socio-economic characteristics. Our findings suggest a trend where older populations are less likely to respond positively, while younger individuals (under 50) are more inclined to contribute. The influence of education and age on WTP is also evident. Educated respondents were more likely to accept the WTP, with those having university degrees willing to pay higher amounts to visit a nature park. Conversely, individuals without higher education were less likely to accept the WTP, with some refusing a 25 € contribution entirely, particularly those with primary or secondary schooling. University-educated respondents demonstrated a strong willingness to pay, with 93% prepared to contribute € 10 or more, including 45% willing to pay € 25 per visit. Only 7% in this category refused to participate. Age was negatively correlated with WTP value, indicating that older individuals were less willing to pay higher amounts. In fact, 60% of those willing to pay were under 35 years old.

### 3.4. Logistic regression

#### 3.4.1. Residents' Willingness to Pay

Building on the initial descriptive analysis, this study aims to identify the variables influencing residents' WTP for the upkeep of their natural environment, particularly the surrounding forest in the Feriana region. Simple descriptive analysis proved to be insufficient for this purpose. Therefore, we employed logistic regression to define a Logit model and gain a deeper understanding of these variables.

The results revealed an average WTP of € 32.6 per household, with an average household composition of 4.8 people and 2.1 children. Notably, residents showed a preference for voluntary bi-monthly payments. Most household heads fell into the low- to medium-income category, with over 60% earning less than € 600 per month. Interestingly, the logistic regression analysis indicated that most respondents favored the development of a leisure park in the surrounding natural areas, with an average preference score of 1.9.

*Educational Level:* Figure 2 reveals a clear trend: individuals with higher education levels (approaching a value of 1) tend to have a higher maximum annual willingness to pay (WTP). This positive correlation is statistically significant ($R^2 = 0.458$).

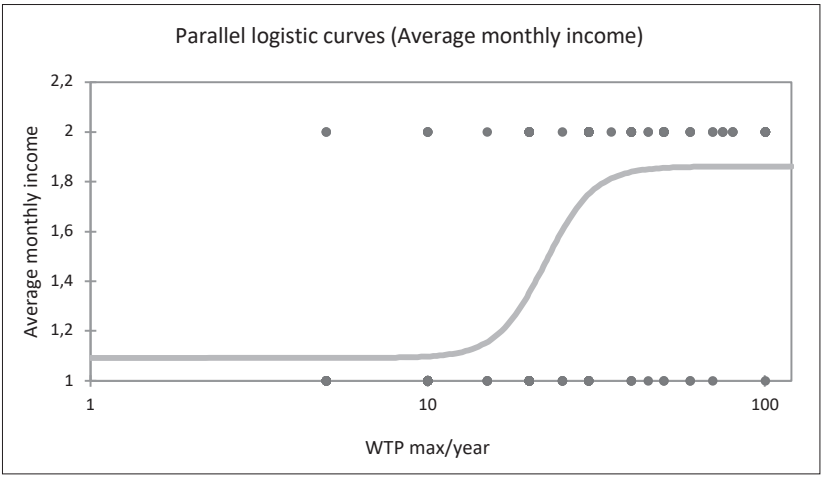

Figure 3 - Effect of the 'average monthly income' variable on the value of WTP.

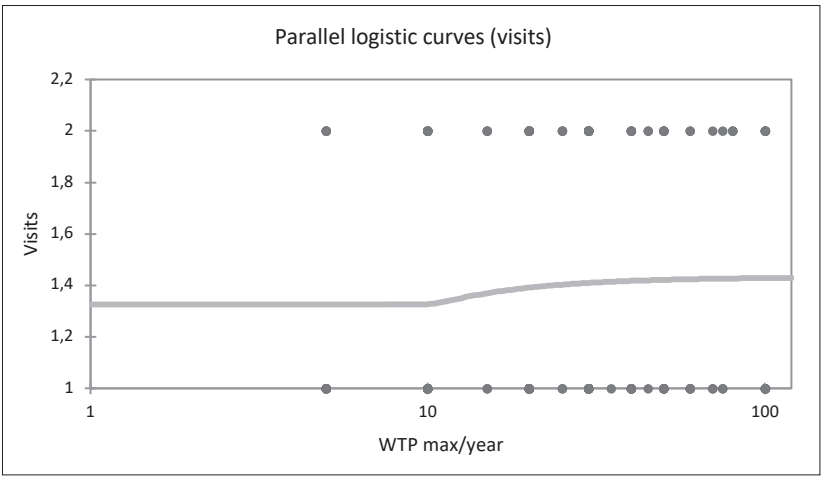

Figure 4 - Effect of the visit variable on the WTP value.

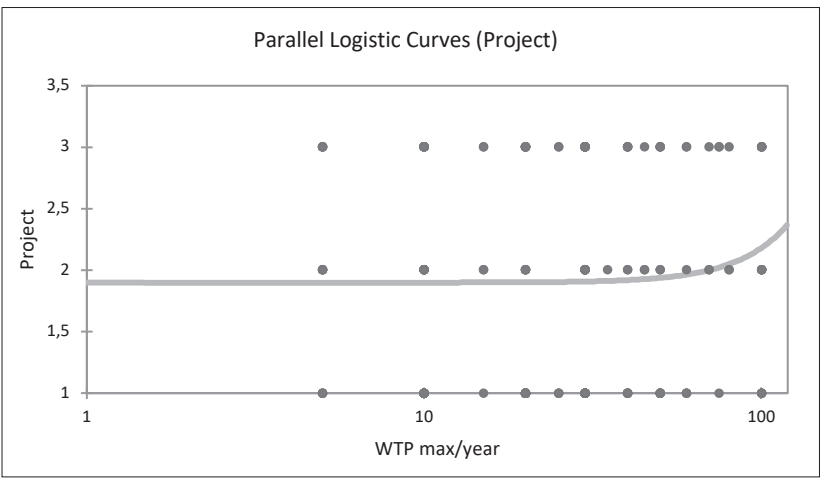

Figure 5 - Effect of the variable 'project type' on the value of WTP.

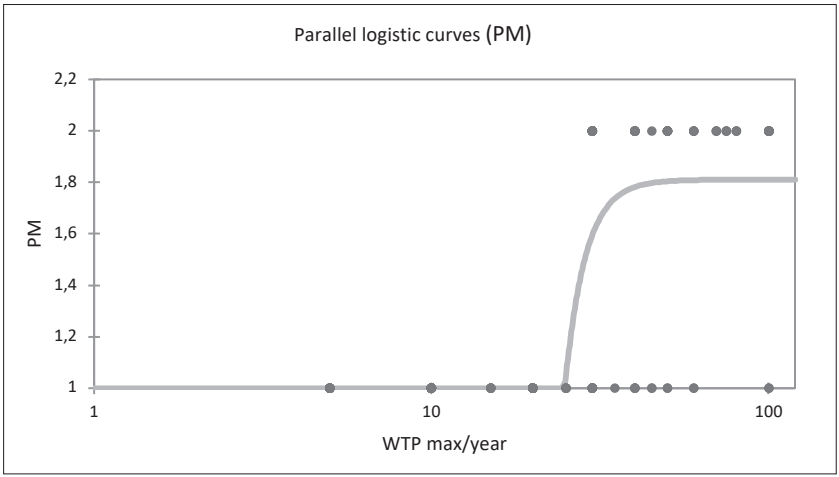

Figure 6 - Effect of the 'Payment method' variable on the WTP value.

*Income*: Figure 3 suggests a positive but weak relationship between WTP and average monthly income. As income increases towards a value of 2 (indicating income above €600), there's a slight increase in WTP. However, the $R^2$ value of 0.028 indicates this correlation is not very strong.

*Frequency of Visits:* Similar to education level, Figure 4 shows a positive correlation between visiting the natural areas and WTP. People who visit more frequently (visit variable closer to 1) are more likely to express a higher WTP. However, the $R^2$ of 0.005 suggests this effect is weakly significant.

*Project Type:* Development appears to be a stronger incentive for higher WTP compared to conservation (Figure 5). This relationship is also weakly significant ($R^2 = 0.005$).

*Payment Method:* The payment method has the most significant impact on WTP ($R^2 = 0.639$). Figure 6 demonstrates that WTP is consistently above €10 and reaches its maximum (€100) when the payment mode tends towards 2 (annual tax).

*Other Variables:* While most variables showed positive correlations with WTP, Figure 7 indicates that educational level and household size might be negatively correlated. In logistic regression, a positive coefficient for a dichotomous variable means those with modality 1 (e.g., visitors) are more likely to pay than those with modality 0 (non-visitors). The opposite applies to negative coefficients. For continuous variables (like income), a positive coefficient suggests that WTP increases as the variable increases (Saadaoui, 2013).

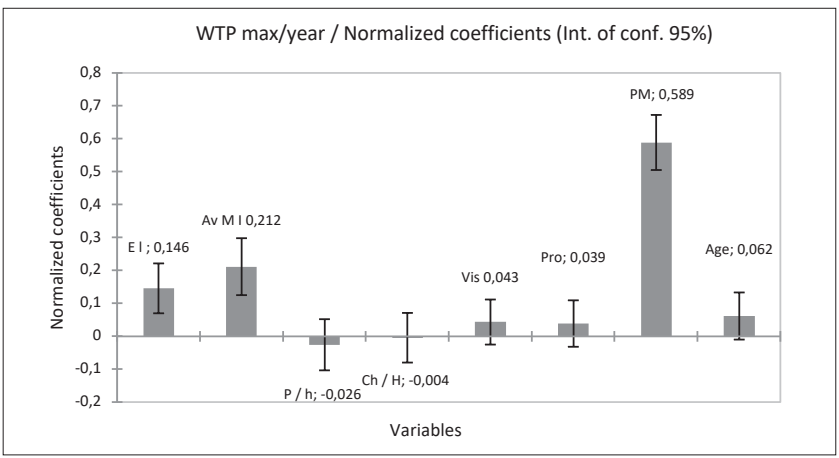

Figure 7 - WTP/Normalized coefficient.

Table 1 - Correlation between descriptive variables and WTP.

| Variables | Educ Level | Monthly Income | Person / House | Child / House | Visits | Project | Payment Mode | Age |
|---|---|---|---|---|---|---|---|---|
| WTP max/ Year | 0,439 | 0,612 | -0,128 | -0,100 | 0,055 | 0,063 | 0,769 | 0,101 |

*Source: XLSTAT 2023 statistical analyzes based on survey data.*

Table 2 - Correlation matrix between the explanatory variables and WTP.

| Variable | Gender | Age | Edu Level | Groupe size | Length stay |
|---|---|---|---|---|---|
| WTP | -0,115 | -0,137 | 0,571 | 0,099 | 0,577 |

*Source: XLSTAT 2023 statistical analyzes based on survey data.*

Analysis of the correlation between descriptive variables and WTP revealed positive correlations for factors including education level, average monthly income, frequency of visits to the natural area, project type (development vs. conservation), preferred payment method, and respondent age. The payment method variable displayed a particularly strong positive correlation with WTP.

Furthermore, Table 1 indicates that average monthly income and preferred payment method have the most significant effects on the monetary value of WTP among the explanatory variables examined.

### 3.4.2. *Tourists' Willingness to Pay*

The second part of the analysis examined tourists' and foreign passengers' WTP for access to a potential natural leisure park. The average WTP estimated by the logistic regression model was € 5450, but this masked a wider range of individual responses. While around 5% of respondents refused to pay entirely, the majority (46%) were willing to pay a more modest entrance fee of € 2.50. An additional 22% and 10% were prepared to pay € 5 and € 20 respectively.

Table 2 (in the referenced table) details the correlations between explanatory variables and the dependent variable, WTP Interestingly, gender and age were negatively correlated with WTP, suggesting a potential preference for the park among younger visitors and those traveling in larger groups. The analysis (Figure 8) also revealed positive correlations between:

*Group Size*: Larger groups were more likely to pay to visit the park.

*Education Level*: Higher education was associated with a greater WTP.

*Length of Stay*: Tourists staying longer in the region were more likely to pay for park access.

These findings suggest that targeting park marketing and development towards younger visitors, larger travel groups, and those with higher education levels could be an effective strategy.

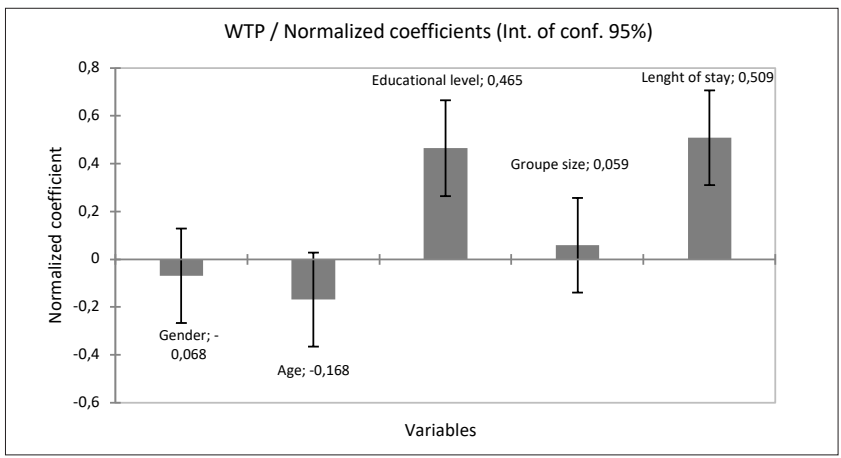

Figure 8 - Descriptive variable correlation /WTP.

Table 3 - Normalized Coefficients (WTP).

| Logit estimates | | | Observations = 50 | | |
| --- | --- | --- | --- | --- | --- |
| Source | Value | t | Pr > \|t\| | Lower Bound (95%) | Upper Bound (95%) |
| Gender | -0,068 | -0,698 | 0,489 | -0,266 | 0,129 |
| Age | -0,168 | -1,723 | 0,092 | -0,365 | 0,029 |
| Educational level | 0,465 | 4,673 | < 0,0001 | 0,264 | 0,665 |
| Family members | 0,059 | 0,602 | 0,550 | -0,139 | 0,257 |
| Length of stay | 0,509 | 5,176 | < 0,0001 | 0,311 | 0,707 |

*Educational level and length of stay drive variability in WTP*

Table 3 (refer to table for details) highlights that educational level and length of stay are the most significant explanatory variables influencing the variability of tourists' WTP for park access. In other words, these factors have the strongest impact on how much tourists are willing to pay to visit the park.

## 4. Discussion

The study investigated stakeholder's preferences for the conservation and development of the Tunisian Aleppo pine forests. Both proposed projects, creating a natural leisure park or maintaining a forested area, aimed to revitalize and enhance the landscape. By comparing these options, the trade-offs between development and preservation were explored, and the economic value residents placed on each approach.

The findings revealed a clear preference for the development project, with residents expressing a higher WTP for a park compared to pure conservation efforts. This suggests that residents value the potential economic and social benefits associated with a leisure park, such as increased tourism and recreational opportunities.

Furthermore, the analysis of payment methods yielded interesting insights. Residents showed a stronger preference for voluntary contributions over annual taxes. This might indicate a desire for more control over their financial contributions or a perception that a tax wouldn't directly benefit the park's development.

The significant influence of education level on WTP is noteworthy. Individuals with higher education levels displayed a greater propensity to pay for both conservation and development projects. This potentially reflects a heightened awareness of the environmental benefits associated with protecting natural areas.

Length of stay also emerged as a significant factor for tourists' WTP. Tourists planning longer stays were more willing to pay for park access, suggesting a greater perceived value from the park experience. These findings can inform marketing strategies by targeting park development and promotion towards younger visitors, larger travel groups, and those with higher education levels who plan extended stays in the region.

Overall, this study sheds light on resident preferences and the potential economic viability of a natural leisure park in the Feriana region. The results highlight the importance of considering both conservation and development objectives, along with resident and tourist perspectives, when formulating sustainable environmental management plans.

Future research could explore the specific features and activities most desired by residents and tourists within a potential park to further optimize its development and economic potential.

A similar study was conducted by Daly-Hassan et al. (2015) to evaluate cork oak forests in Tunisia for the purpose of selecting adaptation options for these forests to climate change. Their findings revealed comparable results regarding the importance of the values of goods and services provided for different uses.

Daly-Hassan and Ben Mansoura (2005) recognized the need for changes in forest legislation to improve management flexibility, product trade, and overall forest health. This included incorporating stakeholders, whose active participation is crucial for sustainable forest development. Even today, after two decades, these changes remain a priority, almost an obligation, to conserve this national treasure and protect the rights of future generations.

## 5. Conclusion

This study employed the CVM to assess the value of the Aleppo pine forests of central Tunisia, considering both conservation and development. Residents were presented with scenarios involving either the preservation of non-harvestable forests or the development of a natural leisure park. The CVM approach allowed us to estimate the economic value that both residents and tourists placed on these natural areas.

Logistic regression analysis provided valuable insights into the factors influencing residents' WTP for environmental improvements. Two key variables emerged as significantly positive influences: monthly income and preferred payment method. Residents with higher incomes demonstrated a greater capacity to contribute financially, while voluntary payments were generally preferred over annual taxes. This highlights the importance of considering affordability and payment flexibility when designing conservation or development initiatives.

The findings demonstrate the potential for reconciliation between economic development and ecological sustainability. Residents expressed a clear preference for the development project, suggesting a perceived value associated with the potential economic and social benefits of a leisure park. However, the positive correlations between WTP and educational level for both conservation and development projects underscore the importance of environmental protection for residents. This highlights the need for future projects to strike a balance between development and conservation, ensuring economic growth while maintaining the ecological value of the natural areas.

This study contributes to a limited body of research within our region, particularly in comparison to the dominance of North American studies in the EVRI database. These findings offer valuable insights applicable to other developing countries grappling with similar challenges of balancing economic development with environmental protection. Active resident engagement in decision-making through valuation methods like CVM can lead to sustainable solutions that benefit both the environment and local communities. Future research could delve deeper into specific park features and activities desired by residents and tourists to further optimize park development and its economic potential, ensuring a future where a thriving natural environment coexists with a prosperous local economy.